\documentclass[twocolumn,prl,superscriptaddress,showpacs,preprintnumbers,amsmath,amssymb]{revtex4}

\usepackage{graphicx}
\usepackage{dcolumn}
\usepackage{bm}

\begin{document}
\title{Phase coexistence of cluster crystals: beyond the Gibbs phase rule}
\author{Bianca~M.~Mladek}\thanks{These authors contributed equally to this work.}\affiliation{Center for Computational Materials Science and Institut f\"ur Theoretische Physik, Technische Universit\"at Wien, Wiedner Hauptstrasse 8-10, 1040
Wien, Austria}
\author{Patrick~Charbonneau}\thanks{These authors contributed equally to this work.}\affiliation{FOM Institute for Atomic and Molecular Physics, Kruislaan 407, 1098 SJ Amsterdam, The Netherlands}
\author{Daan~Frenkel} \email{frenkel@amolf.nl}\affiliation{FOM Institute for Atomic and Molecular Physics, Kruislaan 407, 1098 SJ Amsterdam, The Netherlands}
\date{\today}
\begin{abstract}
We report a study of the phase behavior of multiple-occupancy crystals through simulation.  We argue that in order to reproduce the equilibrium behavior of such crystals it is essential to treat the number of lattice sites as a constraining thermodynamic variable. The resulting free-energy calculations thus differ considerably from schemes used for single-occupancy lattices. Using our approach, we obtain the phase diagram and the bulk modulus for a generalized exponential model that forms cluster crystals at high densities. We compare the simulation results with existing theoretical predictions. We also identify two types of density fluctuations that can lead to two sound modes and evaluate the corresponding elastic constants.
\end{abstract}
 \maketitle

At finite temperatures all crystals contain point defects. This means that the ratio between $N$, the number of particles, and $N_c$, the number of unit cells is not fixed by geometry. In the language of Ref.~\cite{swope:1992}, $N_c$ is a ``constrained'' thermodynamic variable. A general variation of the Helmholtz free energy for a one-component crystal can be written as:
\begin{equation}
dF = -S dT - PdV + \mu dN + \mu_c dN_c,
\end{equation}
where $S$ is the entropy, $T$ the absolute temperature, $P$ the pressure, $V$ the volume, $\mu$ the chemical potential of the constituent particles, and $\mu_c$ the ``cell chemical potential'' conjugate to the number of unit cells. If $N_c$ is free to change, it will take on a value such that $\mu_c=0$, hence its value is a function of $N$, $V$, and $T$. In simple crystals, the equilibrium concentration of point defects is usually so low that their effect on the phase behavior is negligible~\cite{monson:2000,Pronk:2001}. For instance, at melting  the chemical potential of a hard-sphere crystal with vacancies roughly differs by as little as $10^{-3}k_BT$ from that of a defect-free crystal for which $N_c=N$~\cite{Pronk:2001}.

Interestingly, the situation is dramatically different in ``cluster
crystals''~\cite{likos:2001,mladek:2006,likos:2007,fragner:2007,fragner:2007b,mladek:2007}.
These unusual crystalline materials can have a number of particles per lattice site much larger than one. Such solids form in systems of particles that interact via a bounded, short-ranged and purely repulsive pair potential whose Fourier transform has negative regions. The effect of $\mu_c$ on the phase behavior then becomes all important, which has profound consequences for the numerical study of their phase transitions. The reason is that in almost all simulations involving crystals, the average number of particles per unit cell is fixed at the outset of the simulation. After that, a change in the density $\rho\equiv N/V$ of the system may still change $P$ and $\mu$ but, as the ratio $N/N_c$ is fixed, $\mu_c$ will in general not be zero. Hence, conventional simulations do {\em not} probe the lowest free-energy state of the crystal. At constant $P$ and $T$, a small variation in Gibbs free energy $G\equiv F+PV$ is of the form $\mu dN+\mu_c
dN_c$. If we fix the ratio $n_c\equiv N/N_c$, then both $\mu$ and $\mu_c$ are constant, so we can integrate to obtain:
\begin{equation}
G= N\mu+N_c\mu_c
\label{eq:G_muc}
\end{equation}
and hence
\begin{equation}
N_c\mu_c=F+PV-\mu N.
\label{eq:F_muc}
\end{equation}
For a given $N$, $V$, $T$, and $N_c$, we can use Monte Carlo (MC) simulations to compute $F$, $P$, and $\mu$~\cite{frenkel:2002}.
As all quantities on the right-hand side of Eq.~\ref{eq:F_muc} can be determined numerically, whilst $N_c$ is known, we can also compute $\mu_c$. This is important because the condition for phase coexistence involving cluster crystals requires equality of $\mu$, $P$, and $T$ in the coexisting phases {\em and} of $\mu_c=0$ in all crystalline phases. This latter condition is not normally considered in the discussion of the Gibbs phase rule. However, in his original formulation, Gibbs does allow for the possible existence of other thermodynamic ``fields'' in addition to $\mu$, $P$ and $T$~\cite{gibbs:1876}.

As an application of this approach we consider the numerical determination of the phase diagram for the generalized exponential model (GEM-n) $\Phi(r_{ij})=\varepsilon e^{-(r_{ij}/\sigma)^n}$ with $n=4$, where $\varepsilon$ and $\sigma$ determine respectively the energy and the length scales. For convenience, we set them to unity and consider only reduced units from this point forward. For $n>2$, this system is known to form cluster solids at high densities~\cite{likos:2001,mladek:2006,fragner:2007b,moreno:2007}. Its phase diagram is known qualitatively, but not quantitatively: at high $T$, the fluid first freezes into a multiply-occupied BCC phase that transforms into a multiply-occupied FCC phase at higher densities. By contrast, upon compression at low $T$, the system undergoes a ``normal'' freezing transition to a single-occupancy FCC crystal; clustering only sets in upon further compression of the solid.

We perform discretized-space constant-NVT MC simulations~\cite{panagiotopoulos:2000,fragner:2007} for 2000-5000 particles in the temperature regime where multiple occupancy of the crystal lattice sites is expected. We determine the value of $\rho_c\equiv N_c/V$ such that $\mu_c=0$ for every ($\rho$, $T$) point by starting the fixed $N_c$ simulations with a reasonable guess for $N$ and $V$ and iterating until the correct values of $N$ and $V$ are located. Via the common tangent construction, we obtain the coexistence densities from the resulting free energy curves. We note that, because the number of particles per lattice site is free to fluctuate, we cannot use the Einstein-crystal method to compute the free energy of the solid~\cite{frenkel:1984,frenkel:2002}. Rather, we perform a thermodynamic integration from a reference state of ideal-gas particles that move in potential wells centered around the lattice sites~\cite{mladek:2007b}. We also note that for particles that form cluster solids, the Widom particle-insertion method provides an efficient tool to determine the chemical potential, even in the dense solid~\cite{frenkel:2002,widom:1963}. One could even think of performing a kind of Gibbs-ensemble simulation where two systems exchange both particles and volume~\cite{panagiotopoulos:1992}. However, such a simulation would {\em not} locate the correct coexistence point, precisely because the Gibbs-ensemble method does not impose the condition $\mu_c=0$ in all solid phases. That is why we have to follow the rather elaborate route via Eq.~\ref{eq:F_muc} to locate the points where the different phases coexist.

\begin{figure}
\center{\includegraphics[width=\columnwidth]{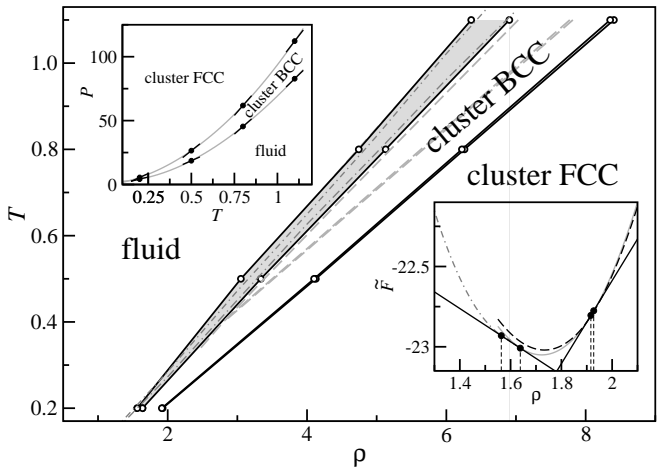}}
\caption[GEM-4 phase diagram]{Comparison of the DFT~\cite{mladek:2006} (dashed lines) and SAMC~\cite{fragner:2007} (dash-dotted lines) $T$-$\rho$ phase diagrams with the simulation results (points) for the GEM-4 model. The gray zone highlights the phase coexistence region. Upper inset: $P$-$T$ simulation phase diagram (points) with the corresponding Clausius-Clapeyron tangents to the coexistence curve (black lines). The solid curves are guides for the eye. Lower inset: shifted free energy curves $\tilde{F}\equiv F/VT-26\rho$ (to enhance visibility) at $T=0.2$ for the liquid (dash-dotted gray line), BCC crystal (solid gray line), and FCC crystal (dashed black line), along with the common tangent construction (solid black lines) and the coexistence densities (dots and drop-down lines).} \label{fig:pd}
\end{figure}

In Fig.~\ref{fig:pd} we compare the $T$-$\rho$ phase diagram for the GEM-4 model with the corresponding density-functional theory (DFT) predictions of Ref.~\cite{mladek:2006}. In the same figure, we also show the estimate of the freezing transition based on the
results of simulated annealing MC (SAMC)~\cite{fragner:2007,fragner:2007b}. For the liquid-BCC transition, the liquidus line predicted by DFT is indistinguishable from the ``exact'' simulation results and the solid line is only slightly off.  The SAMC results, although qualitatively correct, predict a liquid-BCC density gap that is too narrow. This is probably due to finite-size artefacts in Refs.~\cite{fragner:2007,fragner:2007b}. A crucial test of the accuracy of DFT is the prediction of the location of the BCC-FCC transition. As the densities and free energies of these two phases are very close (Fig.~\ref{fig:pd} lower inset), minor inaccuracies in the theory should have a noticeable effect on the prediction of the transition point. Indeed, we find that even though the DFT free-energy predictions are only off by a small amount (not shown), 
the location of the phase transition is shifted by roughly $10\%$ in $\rho$. The $P$-$T$ phase diagram is shown in the upper inset, where the various state points are accompanied by tangents to the coexistence curves obtained from the simulation free and internal energy results by use of the Clausius-Clapeyron relation~\cite{frenkel:2002}. Inspection of the diagram suggests a liquid-BCC-FCC triple point around $T_t\approx0.15$, which is much lower than the DFT prediction of $T_t \approx 0.4$~\cite{mladek:2006}. The dramatic shift follows from the small difference between the slopes of the liquid-BCC and the BCC-FCC coexistence curves, leaving the location of the triple point very sensitive to any modification of the latter.

\begin{figure}
\center{\includegraphics[width=\columnwidth]{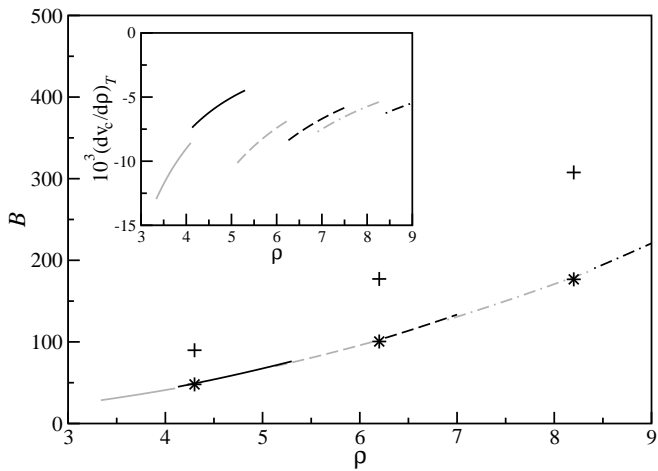}}
\caption[Bulk modulus in GEM-4]{Bulk modulus results from direct differentiation of the free energy for three different temperatures in the stable crystal structures [BCC (gray) and FCC (black) at $T=0.5$ (solid line), $T=0.8$ (dashed lined), and $T=1.1$ (dash-dotted line)], along with the values at three state points (stars) computed using Eq.~\ref{eq:bulkmod}. The virial contribution to $B$ is shown for reference (crosses). The breakdown of $B$ is also given in Table~\ref{tbl:elastic}. Inset:
variation of the lattice volume with density in equilibrium for these same systems. DFT presents this quantity as zero.}\label{fig:bulkmod}
\end{figure}

It is interesting to understand the reason for the failure of DFT to predict the location of the solid-solid transition.  One of the core consequences of the DFT treatment of Ref.~\cite{mladek:2006} is that the volume of the unit cell in a cluster crystal $v_c \equiv V/N_c$ is independent of density, so that the average site occupation $n_c \propto \rho$. This feature is known to break down in low-density
crystals~\cite{fragner:2007} and is also found here to be slightly inaccurate at intermediate temperatures and densities. Though the linear relationship holds, the proportionality is shifted by a constant. As can be gathered from the inset of Fig.~\ref{fig:bulkmod}, this leads to a non-zero value of $(\partial v_c/\partial \rho)_{T}$ for equilibrium states, though the effect vanishes with increasing $T$ and $\rho$. This suggests that the DFT approximation is asymptotically valid. At intermediate densities, this correction though small might be sufficient to explain the discrepancy between the DFT and the numerical results.

Since thermodynamic equilibrium is only obtained when $\mu_c=0$, it would appear that once the equilibrium points are found, all references to the artificial $\mu_c$ field can be disregarded.  Yet, for quantities involving the second derivative of the constrained free energy, such as the bulk modulus $B=V \left(\frac{\partial^2 F}{\partial V^2} \right)_{N,T}$, it cannot be neglected unless one already has the complete {\it equilibrium} free energy curve at hand. In simulations of single-occupancy crystals, $B_{\rm vir}=-V\left(\frac{\partial P}{\partial V}\right)_{N,T,N_c}$ can be computed directly for a given state point through an approach similar to the virial calculation of $P$~\cite{rowlinson:1959,cheung:1977}. For cluster crystals, however, the artificial system conditions further modify the bulk modulus as
\begin{eqnarray}
B=B_{\rm vir}-\frac{\rho^2}{n_c}\left(\frac{\partial \mu_c}{\partial
\rho}\right)_{T,n_c}\left[1+\frac{\rho}{v_c}\left(\frac{\partial
v_c}{\partial \rho}\right)_{T,\mu_c=0}\right],
\label{eq:bulkmod}
\end{eqnarray}
where the partial derivatives are evaluated around an equilibrium state point. The virial contribution corresponds to a quenched system where particle rearrangements are not possible, so it is an upper bound to $B\equiv B_{\rm vir}-B_{\rm corr}$. The results for different state points are compared in Fig.~\ref{fig:bulkmod} to the values obtained by direct numerical differentiation of the equilibrium free energy curves. Remarkable agreement is obtained between the two approaches. Also, far from negligible, $B_{\rm corr}$ results in a $B$ about $40\%$ smaller than $B_{\rm vir}$, as can also be gathered from Table~\ref{tbl:elastic}. The leading term to the correction, the change in $\mu_c$ with density, suggests that deletion of lattice sites weakens the system's response to compression. The changes in lattice site occupancy permitted by particle overlap thus increase the crystal compressibility compared to simple affine transformations. Generally, this still translates into an increase of $B$ with density. Note also that the temperature dependence is rather weak. For $T=0.5-1.1$, the curves in Fig.~\ref{fig:bulkmod} appear to collapse onto a master function, which suggests that entropic effects have little impact in this regime.

\begin{figure}
\center{\includegraphics[width=\columnwidth]{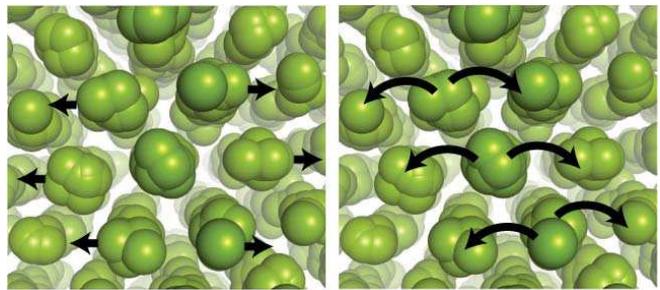}}
\caption[MOC sound modes]{(Color online) Density fluctuations in multiple occupancy crystals either stem from (left) fluctuations in the unit cell volume or from (right) rearrangements in particle distributions between lattice sites.} \label{fig:sound}
\end{figure}

If sound waves have a period shorter than the time it takes for particles to redistribute between unit cells, then we can distinguish between two different sound modes in cluster solids. Density fluctuations stem either from changes to the unit cell volume $v_c$ at fixed cluster occupancy $n_c$ or from fluctuations in $n_c$ at fixed $v_c$, as schematized in Fig.~\ref{fig:sound}. At constant $T$ and $V$, the free energy density fluctuations $\Delta f\equiv \Delta F/V$ are then
\begin{equation}
\Delta f=c_{11} (\rho_c\Delta n_c)^2
+c_{22} (n_c\Delta \rho_c)^2+c_{12} \rho_c n_c\Delta n_c\Delta
\rho_c,
\end{equation}
where the elastic constants
\begin{eqnarray}
c_{11}&=&\frac{1}{2\rho_c}\left(\frac{\partial \mu}{\partial
n_c}\right)_{\rho_c},~~~c_{22}=\frac{1}{2n_c^2}\left(\frac{\partial
\mu_c +\mu n_c}{\partial \rho_c}\right)_{n_c},\nonumber\\
c_{12}&=&\frac{1}{\rho}\left(\frac{\partial \mu
\rho_c}{\partial \rho_c}\right)_{n_c}\nonumber
\end{eqnarray}
can be obtained by numerically differentiating the simulation results.
A change of variables to sound-mode space
$x_\pm=\sqrt{c_{11}}\rho_c\Delta n_c\pm\sqrt{c_{22}}n_c\Delta \rho_c$,
diagonalizes the expression
\begin{equation}
\Delta f=(1+\beta)x_{+}^2+(1-\beta)x_{-}^2
\end{equation}
with coupling constant $\beta\equiv c_{12}/\sqrt{c_{11} c_{22}}$. The elastic coefficients and the coupling constant are presented for three different state points in Table~\ref{tbl:elastic}. Though these results are insufficient to paint the full physical picture, a couple of comments are in order. First, the $c_{11}$ term corresponds to the ``permeation'' of particles and is expected to be heavily damped, while the $c_{22}$ term is the equivalent of a longitudinal sound wave, which will propagate for long wavelengths. Second, for the temperature and density range under study the first constant and the cross term increase only very little with $T$ and $\rho$, while the $c_{22}$ more than doubles. The increased density from one state point to the next is most certainly responsible for that, since higher temperatures would tend instead to facilitate lattice spacing fluctuations for a constant repulsive energy barrier. To the best of our knowledge, no theoretical predictions exist with which to further compare these results.

\begin{table}
  \centering
\begin{tabular}{|c c|c c c|c c c c|}
\hline
\hline
$T$ & $\rho$ & $B$ & $B_{\rm vir}$ & $B_{\rm corr}$ & $c_{11}$ & $c_{12}$ & $c_{22}$ & $\beta$ \\\hline
0.5 & 4.3 & 48.2 & 89.8  & 41.6  & 1.335 & 2.95 & 10.03 & 0.805 \\
0.8 & 6.2 & 100.5 & 177.1 & 76.6  & 1.346 & 3.05 & 15.35 & 0.670 \\
1.1 & 8.2 & 176.6 & 308 & 131.0 & 1.350 & 3.10 & 21.0 & 0.582 \\
\hline
\hline
\end{tabular}
\caption{Bulk modulus decomposition and the sound mode elastic and coupling constants
for three different multiply-occupied crystal state points.}\label{tbl:elastic}
\end{table}

Starting from the formalism developed by Swope and Andersen~\cite{swope:1992}, we have presented how simulations and experiments of multiple-occupancy crystals critically depend on the chemical potential associated with the insertion of a lattice site. Taking this into account within a simulation allows for the precise determination of the equilibrium phase diagram of cluster crystals, which is much more subtle than for the traditional, single-occupancy sort. Also, even though the chemical potential associated with lattice site insertion is strictly zero in equilibrium, its constrained derivatives are not. This has considerable impact on the calculation the bulk modulus and the two sound modes' elastic constants, for example. Departing more drastically from thermodynamics, long-lived non-equilibrium structures of cluster-crystal forming dendrimers might even be observable in rapidly quenched experimental systems, if the resulting ordered solids happen to end up in states with $\mu_c\neq 0$. These metastable crystals would then undergo phase transitions at different state points than those predicted by equilibrium thermodynamics. Finally, the generalization of the free energy calculation methodology presented here has a broader applicability than for multiply-occupied crystals. It would also be the appropriate way to simulate systems with variable occupancy of lattice sites, such a micellar crystals or microphase separated colloids.

We would like to thank D.~Gottwald, G.~Kahl, C.~N.~Likos, M.~Neumann, and B.~Widom for helpful discussions. The work of the FOM Institute is part of the research program of FOM and is made possible by financial support >from the Netherlands Organization for Scientific Research (NWO). Moreover, this work was carried out under the HPC-EUROPA project (RII3-CT-2003-506079), with the support of the European Community - Research Infrastructure Action of the FP6. Computer time at the Dutch center for high-performance computing SARA is gratefully acknowledged.  P.C.~further acknowledges MIF1-CT-2006-040871 (EU) funding and B.M.M. acknowledges financial support by the \"Osterreichische Forschungsfond (FWF) under Project No.~P17823-N08 as well as travel funding by COST-P13 (MOLSIMU) and the \"Osterreichische Forschungsgemeinschaft.


\begin{thebibliography}{19}
\expandafter\ifx\csname natexlab\endcsname\relax\def\natexlab#1{#1}\fi
\expandafter\ifx\csname bibnamefont\endcsname\relax
  \def\bibnamefont#1{#1}\fi
\expandafter\ifx\csname bibfnamefont\endcsname\relax
  \def\bibfnamefont#1{#1}\fi
\expandafter\ifx\csname citenamefont\endcsname\relax
  \def\citenamefont#1{#1}\fi
\expandafter\ifx\csname url\endcsname\relax
  \def\url#1{\texttt{#1}}\fi
\expandafter\ifx\csname urlprefix\endcsname\relax\def\urlprefix{URL }\fi
\providecommand{\bibinfo}[2]{#2}
\providecommand{\eprint}[2][]{\url{#2}}

\bibitem[{\citenamefont{Swope and Andersen}(1992)}]{swope:1992}
\bibinfo{author}{\bibfnamefont{W.~C.} \bibnamefont{Swope}} \bibnamefont{and}
  \bibinfo{author}{\bibfnamefont{H.~C.} \bibnamefont{Andersen}},
  \bibinfo{journal}{Phys. Rev. A} \textbf{\bibinfo{volume}{46}},
  \bibinfo{pages}{4539} (\bibinfo{year}{1992}).

\bibitem[{\citenamefont{Monson and Kofke}(2000)}]{monson:2000}
\bibinfo{author}{\bibfnamefont{P.~A.} \bibnamefont{Monson}} \bibnamefont{and}
  \bibinfo{author}{\bibfnamefont{D.~A.} \bibnamefont{Kofke}}, in
  \emph{\bibinfo{booktitle}{Advances in Chemical Physics}}, edited by
  \bibinfo{editor}{\bibfnamefont{I.}~\bibnamefont{Prigogine}} \bibnamefont{and}
  \bibinfo{editor}{\bibfnamefont{S.~A.} \bibnamefont{Rice}}
  (\bibinfo{publisher}{{John Wiley \& Sons}}, \bibinfo{address}{New York},
  \bibinfo{year}{2000}), vol. \bibinfo{volume}{115} of
  \emph{\bibinfo{series}{Advances in Chemical Physics}}, pp.
  \bibinfo{pages}{113--179}.

\bibitem[{\citenamefont{Pronk and Frenkel}(2001)}]{Pronk:2001}
\bibinfo{author}{\bibfnamefont{S.}~\bibnamefont{Pronk}} \bibnamefont{and}
  \bibinfo{author}{\bibfnamefont{D.}~\bibnamefont{Frenkel}},
  \bibinfo{journal}{J. Phys. Chem. B} \textbf{\bibinfo{volume}{105}},
  \bibinfo{pages}{6722} (\bibinfo{year}{2001}).

\bibitem[{\citenamefont{Likos et~al.}(2001)\citenamefont{Likos, Lang,
  Watzlawek, and L\"owen}}]{likos:2001}
\bibinfo{author}{\bibfnamefont{C.~N.} \bibnamefont{Likos}},
  \bibinfo{author}{\bibfnamefont{A.}~\bibnamefont{Lang}},
  \bibinfo{author}{\bibfnamefont{M.}~\bibnamefont{Watzlawek}},
  \bibnamefont{and} \bibinfo{author}{\bibfnamefont{H.}~\bibnamefont{L\"owen}},
  \bibinfo{journal}{Phys. Rev. E} \textbf{\bibinfo{volume}{63}},
  \bibinfo{pages}{031206} (\bibinfo{year}{2001}).

\bibitem[{\citenamefont{Mladek et~al.}(2006)\citenamefont{Mladek, Gottwald,
  Kahl, Neumann, and Likos}}]{mladek:2006}
\bibinfo{author}{\bibfnamefont{B.~M.} \bibnamefont{Mladek}},
  \bibinfo{author}{\bibfnamefont{D.}~\bibnamefont{Gottwald}},
  \bibinfo{author}{\bibfnamefont{G.}~\bibnamefont{Kahl}},
  \bibinfo{author}{\bibfnamefont{M.}~\bibnamefont{Neumann}}, \bibnamefont{and}
  \bibinfo{author}{\bibfnamefont{C.~N.} \bibnamefont{Likos}},
  \bibinfo{journal}{Phys. Rev. Lett.} \textbf{\bibinfo{volume}{96}},
  \bibinfo{pages}{045701} (\bibinfo{year}{2006}).

\bibitem[{\citenamefont{Likos et~al.}(2007)\citenamefont{Likos, Mladek,
  Gottwald, and Kahl}}]{likos:2007}
\bibinfo{author}{\bibfnamefont{C.~N.} \bibnamefont{Likos}},
  \bibinfo{author}{\bibfnamefont{B.~M.} \bibnamefont{Mladek}},
  \bibinfo{author}{\bibfnamefont{D.}~\bibnamefont{Gottwald}}, \bibnamefont{and}
  \bibinfo{author}{\bibfnamefont{G.}~\bibnamefont{Kahl}}, \bibinfo{journal}{J.
  Chem. Phys.} \textbf{\bibinfo{volume}{126}}, \bibinfo{pages}{224502}
  (\bibinfo{year}{2007}).

\bibitem[{\citenamefont{Fragner}(2007{\natexlab{a}})}]{fragner:2007}
\bibinfo{author}{\bibfnamefont{H.}~\bibnamefont{Fragner}},
  \bibinfo{journal}{Phys. Rev. E} \textbf{\bibinfo{volume}{75}},
  \bibinfo{pages}{061402} (\bibinfo{year}{2007}{\natexlab{a}}).

\bibitem[{\citenamefont{Fragner}(2007{\natexlab{b}})}]{fragner:2007b}
\bibinfo{author}{\bibfnamefont{H.}~\bibnamefont{Fragner}},
  \bibinfo{journal}{Comput. Phys. Commun.} \textbf{\bibinfo{volume}{176}},
  \bibinfo{pages}{327} (\bibinfo{year}{2007}{\natexlab{b}}).

\bibitem[{\citenamefont{Mladek et~al.}(2007{\natexlab{a}})\citenamefont{Mladek,
  Gottwald, Kahl, Neumann, and Likos}}]{mladek:2007}
\bibinfo{author}{\bibfnamefont{B.~M.} \bibnamefont{Mladek}},
  \bibinfo{author}{\bibfnamefont{D.}~\bibnamefont{Gottwald}},
  \bibinfo{author}{\bibfnamefont{G.}~\bibnamefont{Kahl}},
  \bibinfo{author}{\bibfnamefont{M.}~\bibnamefont{Neumann}}, \bibnamefont{and}
  \bibinfo{author}{\bibfnamefont{C.~N.} \bibnamefont{Likos}},
  \bibinfo{journal}{J. Phys. Chem. B}  (\bibinfo{year}{2007}{\natexlab{a}}),
  \bibinfo{note}{(in press)}.

\bibitem[{\citenamefont{Frenkel and Smit}(2002)}]{frenkel:2002}
\bibinfo{author}{\bibfnamefont{D.}~\bibnamefont{Frenkel}} \bibnamefont{and}
  \bibinfo{author}{\bibfnamefont{B.}~\bibnamefont{Smit}},
  \emph{\bibinfo{title}{Understanding Molecular Simulation}},
  vol.~\bibinfo{volume}{1} of \emph{\bibinfo{series}{Computational Science
  Series}} (\bibinfo{publisher}{Academic Press}, \bibinfo{address}{San Diego},
  \bibinfo{year}{2002}).

\bibitem[{\citenamefont{Gibbs}(1993)}]{gibbs:1876}
\bibinfo{author}{\bibfnamefont{J.~W.} \bibnamefont{Gibbs}},
  \emph{\bibinfo{title}{The Scientific Papers of J. Willard Gibbs, Volume 1:
  Thermodynamics}} (\bibinfo{publisher}{Ox Bow Press},
  \bibinfo{address}{Woodbridge}, \bibinfo{year}{1993}).

\bibitem[{\citenamefont{Moreno and Likos}(2007)}]{moreno:2007}
\bibinfo{author}{\bibfnamefont{A.~J.} \bibnamefont{Moreno}} \bibnamefont{and}
  \bibinfo{author}{\bibfnamefont{C.~N.} \bibnamefont{Likos}},
  \bibinfo{journal}{Phys. Rev. Lett.}  (\bibinfo{year}{2007}),
  \bibinfo{note}{(in press), eprint arXiv:cond-mat/0706.0659v1}.

\bibitem[{\citenamefont{Panagiotopoulos}(2000)}]{panagiotopoulos:2000}
\bibinfo{author}{\bibfnamefont{A.~Z.} \bibnamefont{Panagiotopoulos}},
  \bibinfo{journal}{J. Chem. Phys.} \textbf{\bibinfo{volume}{112}},
  \bibinfo{pages}{7132} (\bibinfo{year}{2000}).

\bibitem[{\citenamefont{Frenkel and Ladd}(1984)}]{frenkel:1984}
\bibinfo{author}{\bibfnamefont{D.}~\bibnamefont{Frenkel}} \bibnamefont{and}
  \bibinfo{author}{\bibfnamefont{A.~J.~C.} \bibnamefont{Ladd}},
  \bibinfo{journal}{J. Chem. Phys.} \textbf{\bibinfo{volume}{81}},
  \bibinfo{pages}{3188} (\bibinfo{year}{1984}).

\bibitem[{\citenamefont{Mladek et~al.}(2007{\natexlab{b}})\citenamefont{Mladek,
  Charbonneau, and Frenkel}}]{mladek:2007b}
\bibinfo{author}{\bibfnamefont{B.~M.} \bibnamefont{Mladek}},
  \bibinfo{author}{\bibfnamefont{P.}~\bibnamefont{Charbonneau}},
  \bibnamefont{and} \bibinfo{author}{\bibfnamefont{D.}~\bibnamefont{Frenkel}}
  (\bibinfo{year}{2007}{\natexlab{b}}), \bibinfo{note}{(unpublished)}.

\bibitem[{\citenamefont{Widom}(1963)}]{widom:1963}
\bibinfo{author}{\bibfnamefont{B.}~\bibnamefont{Widom}}, \bibinfo{journal}{J.
  Chem. Phys.} \textbf{\bibinfo{volume}{39}}, \bibinfo{pages}{2808}
  (\bibinfo{year}{1963}).

\bibitem[{\citenamefont{Panagiotopoulos}(1992)}]{panagiotopoulos:1992}
\bibinfo{author}{\bibfnamefont{A.~Z.} \bibnamefont{Panagiotopoulos}},
  \bibinfo{journal}{Mol. Simul.} \textbf{\bibinfo{volume}{9}},
  \bibinfo{pages}{1} (\bibinfo{year}{1992}).

\bibitem[{\citenamefont{Rowlinson}(1959)}]{rowlinson:1959}
\bibinfo{author}{\bibfnamefont{J.}~\bibnamefont{Rowlinson}},
  \emph{\bibinfo{title}{Liquids and liquid mixtures}}, Modern aspects series of
  chemistry (\bibinfo{publisher}{Butterworths Scientific Publications},
  \bibinfo{address}{London}, \bibinfo{year}{1959}).

\bibitem[{\citenamefont{Cheung}(1977)}]{cheung:1977}
\bibinfo{author}{\bibfnamefont{P.~S.~Y.} \bibnamefont{Cheung}},
  \bibinfo{journal}{Mol. Phys.} \textbf{\bibinfo{volume}{33}},
  \bibinfo{pages}{519} (\bibinfo{year}{1977}).

\end{thebibliography}
\end{document}